\documentclass[a4paper,11pt]{article}
\usepackage{jheppub} 
\usepackage{epsfig}
\usepackage{graphicx}
\usepackage[T1]{fontenc}
\def\be{\begin{equation}}
\def\ee{\end{equation}}
\def\bea{\begin{array}}
\def\eea{\end{array}}
\def\beqa{\begin{eqnarray}}
\def\eeqa{\end{eqnarray}}
\def\beqas{\begin{eqnarray*}}
\def\eeqas{\end{eqnarray*}}

\def\bp{\begin{picture}}
\def\ep{\end{picture}}
\def\bc{\begin{center}}
\def\ec{\end{center}}
\def\bfig{\begin{figure}}
\def\efig{\end{figure}}

\def\bit{\begin{itemize}}
\def\eit{\end{itemize}}
\def\nn{\nonumber}
\def\f{\frac}

\def\[{\left[}
\def\]{\right]}
\def\({\left(}
\def\){\right)}

\def\..{\left.}
\def\.{\right.}
\def\tl{\tilde}
\def\ra{\rightarrow}

\def\tm{\times}

\def\al{\alpha}

\def\ep{\epsilon}

\def\ga{\gamma}
\def\pa{\partial}
\def\pr{\prime}

\title{\boldmath Relativistic origin of Hertz-form and extended Hertz-form equations for
Maxwell theory of electromagnetism}
\author[a,1]{Fei Wang\note{Corresponding author.}}
\author[b,c]{Jin Min Yang}
\affiliation[a]{Department of Physics, Zhengzhou University, No. 100 Science Avenue, ZhengZhou 450001, P. R. China}
\affiliation[b]{CAS Key Laboratory of Theoretical Physics, Institute of Theoretical Physics, Chinese Academy of
Sciences, Beijing 100190, P. R. China}
\affiliation[c]{School of Physical Sciences, University of Chinese Academy of Sciences,  Beijing 100049, P. R. China }
\emailAdd{feiwang@zzu.edu.cn,jmyang@itp.ac.cn}

\abstract{We show explicitly that the Hertz-form Maxwell's equations and their extensions
can be obtained from the non-relativistic expansion of Lorentz transformation of Maxwell's equations.
The explicit expression for the parameter $\alpha$ in the extended Hertz-form equations can be derived
from such a non-relativistic expansion. The extended Hertz-form equations,
which do not preserve Galilean invariance, origin from Lorentz transformation of Maxwell's equations and differ from the Galilean-transformed Maxwell equations (the original Hertz equations) by the relative sign differences between the two $\alpha$ terms etc.
Especially, the $\alpha$ parameter is of relativistic origin. The superluminal behavior illustrated by
the D'Alembert equation from the extended Hertz-form equations should be removed by including
all subleading contributions in the $v/c$ expansion, although such a superluminal behavior will not occur in the vacuum because $\al=0$. In the case that the electromagnetic
field is a background field, we need not worry about the apparent superluminal behavior of the
D'Alembert equation. We should note that in the Hertz form and extended Hertz form equations,
the electromagnetic fields should take the forms
$ \vec{\mathcal{E}}(x)=\vec{E}(\Lambda^{-1}x)$ and $ \vec{\mathcal{B}}(x)=\vec{B}(\Lambda^{-1}x)$
while the derivatives in the equations are taken with respect to $x$.
Such a choice of description for the fields is different from the ordinary one with
$\vec{E}(x)$ and $\vec{B}(x)$, which are well known to satisfy the ordinary Maxwell's equations.
The descriptions of electromagnetic phenomena using the function set $\{\vec{\mathcal{E}}(x),\vec{\mathcal{B}}(x)\}$
and the function set $(\vec{E}(x),\vec{B}(x))$ are equivalent, with
the $\{\vec{\mathcal{E}}(x),\vec{\mathcal{B}}(x)\}$ description satisfying
the extended Hertz-form Maxwell's equations in the low speed approximation. The solution of
(extended) Hertz-form Maxwell's equations describe the traveling wave form electromagnetic field.
}
\begin{document}
\maketitle
\flushbottom

\section{Introduction}
Maxwell's equations, which can successfully describe the classical electromagnetic phenomena, play an important rule in
fundamental science and practical technologies. There exist various extensions to Maxwell's equations, mostly by adding
the magnetic monopole related terms in the framework of quantum field theory.
In the framework of classical limit,
the Hertz-form equations~\cite{hertz}, which amend the ordinary Maxwell equations by several terms,
apparently preserve Galilean invariance.
 As the Maxwell's equations should satisfy Lorentz invariance, the Hertz-form Maxwell's equations seem strange.
So, it is interesting to revisit Hertz-form equations to figure out if it is possible
to derive them from the non-relativistic approximation of Lorentz transformation for electromagnetic fields.

However, the Hertz-form equations do not agree with the experimental data on the movement of dielectrics
in electromagnetic
fields ~\cite{pauli}. In order to satisfy the experimental data, in ~\cite{ru} an additional factor
$\alpha=(\mu_r\epsilon_r-1)/\mu_r\epsilon_r$ is introduced for the terms $\nabla\times(B\times v)$
and $\nabla\times(D\times v)$ appearing in Hertz-form equations of $\nabla\times E $ and  $\nabla\times H $,
with $\epsilon_r$ and $\mu_r$ being the relative permittivity and the relative permeability of the medium,
and $E$, $H$, $B$, $D$ being the vectors of electric and magnetic field strengths, magnetic
induction and electric displacement, respectively. Such extended Hertz-form equations with the parameter $\alpha$,
which lose Galilean invariance, can agree with the experimental data~\cite{ru}. Although the appearance of
the factor $\alpha$ is commented in~\cite{ru}, its origin is not specified. So, it is interesting to figure out
the origin of the $\alpha$ parameter from the first principle.

Note that the recent work in \cite{wzl} proposes to extend Maxwell's equations
with new $P_S$ term ~\cite{wzl2} and velocity-related terms to describe electromagnetic phenomena
in the slow-moving media. Formally, after redefining a new electric
displacement field $D=D^\pr+P_S$, the work of ~\cite{wzl} can be seen to be equivalent to
Hertz-form equations~\cite{hertz}. Maxwell equations in materials have also been studied
intensively in~\cite{lax,bladel}.

In this note, we derive the non-relativistic approximation for Lorentz transformation of Maxwell's equations.
We find that the (extended) Hertz-form equations can be obtained at the (next) leading order approximation
and the explicit form of $\alpha$ can be obtained from the non-relativistic limit. We show that the
appearance of the factor $\alpha$, which is commented in \cite{ru} and violates Galilean transformation,
is the consequence of special relativity. The (extended) Hertz-form Maxwell's equations give an alternative (traveling-wave form) solution for electromagnetic field.

This paper is organized as follows. In Sec.~\ref{sec-2}, we derive explicitly the form of $\alpha$ and
the extended Hertz-form equations from the non-relativistic expansion of Lorentz transformation for
Maxwell's equations. A typical consequence of the extended Hertz-form equations, i.e., the superluminal behavior,
is discussed in Sec.~\ref{sec-3}.
Sec.~\ref{conclusion} contains our conclusions. Some important formulas used in the non-relativistic (low speed) approximation
are given in the appendix~\ref{appendix:2}.

\section{\label{sec-2} Non-relativistic expansion of Lorentz transformation for Maxwell's equations}
The validity of applying special relativity to the description of electromagntic field and its quanta
has been proven to an astonishing accuracy. For example, the theoretical prediction on electron anomalous
magnetic moment by QED can fit the experimental data up to $10^{-10}$ accuracy. So, to describe electromagnetic
phenomena, it is natural to begin
with special relativity. In the following, we try to derive the non-relativistic approximation of Lorentz
transformation for Maxwell's equations to describe the relevant electromagnetic phenomena in slow-moving media.

Lorentz transformation of electromagnetic field is given by
\beqa
F^\pr_{\mu\nu}=\Lambda_\mu^\rho \Lambda_\nu^\sigma F_{\rho\sigma}~,
\eeqa
with
\beqa
F_{\mu\nu}=\pa_\mu A_\nu-\pa_\nu A_\mu~,
\eeqa
which, in components, is given by
\beqa
F_{\mu\nu}=\(\bea{cccc} 0 & B_3&-B_2&-E_1/c\\
                       -B_3&0&B_1&-E_2/c\\
                        B_2&-B_1&0&-E_3/c \\
                        E_1/c & E_2/c & E_3/c &0 \eea\).
\eeqa
We keep the vacuum light speed $c$ in the expressions for later convenience.

In components, we have Lorentz transformation of electromagnetic fields as
\beqa
\vec{E}_{\bot}\ra\ga(\vec{E}+\vec{v}\tm \vec{B}\f{}{})_{\bot}~,
~~~\vec{B}_{\bot} \ra \ga\(\vec{B}-\f{\vec{v}\tm \vec{E}}{c^2}\)_{\bot}~,
\eeqa
and
\beqa
\vec{E}_{\shortparallel}\ra \vec{E}_{\shortparallel}~,~~~\vec{B}_{\shortparallel}\ra \vec{B}_{\shortparallel}~,
\eeqa
with
\beqa
\ga=\f{1}{\sqrt{1-v^2/c^2}}~.
\eeqa
From the transformation law, it is obvious that we only need to discuss the simplest case with $\vec{E}$ and $\vec{B}$
perpendicular to the constant velocity $\vec{v}$ of the reference frame.
The extensions to general cases are straightforward.

In our analysis we consider two  reference frames, one is $\Sigma$ (laboratory) frame with coordinates $\{x\}$,
and the other is $\Sigma'$ (loop-rest) frame with coordinates $\{x^\pr\}$ which is at rest with respect to the medium
and is moving at a constant velocity $\vec{v}$ with respect to $\Sigma$.
Then we have the relations between the fields
\beqa
\vec{E}_{\bot}(x)&\ra&\vec{E}_{\bot}^\pr(x^\pr)=\ga(\vec{E}(x)+\vec{v}\tm \vec{B}(x)\f{}{})_{\bot}~,\nn\\
~~~\vec{B}_{\bot}(x) &\ra&\vec{B}_{\bot}^\pr(x^\pr)=\ga\(\vec{B}(x)-\f{\vec{v}\tm \vec{E}(x)}{c^2}\)_{\bot}~,
\eeqa
and the relation between the coordinates
\beqa
 x'=\Lambda x~,
\eeqa
with $\Lambda$ the Lorentz transformation for the coordinates.
Alternatively, we can write in the forms of $\vec{\mathcal{E}}(x^\pr)$ and $\vec{\mathcal{B}}(x^\pr)$
\beqa
\vec{E}_{\bot}^\pr(x^\pr)&=&\ga(\vec{\mathcal{E}}(x^\pr)+\vec{v}\tm \vec{\mathcal{B}}(x^\pr)\f{}{})_{\bot}~,\nn\\
\vec{B}_{\bot}^\pr(x^\pr)&=&\ga\(\vec{\mathcal{B}}(x^\pr)-\f{\vec{v}\tm \vec{\mathcal{E}}(x^\pr)}{c^2}\)_{\bot}~,
\label{huati-E-B}
\eeqa
where
\beqa
\vec{\mathcal{E}}(x^\pr)\equiv\vec{E}[x(x^\pr)]~,~\vec{\mathcal{B}}(x^\pr)\equiv\vec{B}[x(x^\pr)]~,
\label{function:form}
\eeqa
with
\beqa
x(x')=\Lambda^{-1} x^\pr~,
\eeqa
being the inverse coordinate transformation.
The explicit function forms of $\vec{E}(x)$ and $\vec{\mathcal{E}}(x)$ are different.

In the non-relativistic approximation up to $o(v/c)$, we have the approximate Galilean transformation for coordinates
\beqa
\vec{x}^\pr=\vec{x}-\vec{v} t~,~~t^\pr=t-\f{\vec{x}\cdot \vec{v}}{c^2},
\eeqa
and thus we have
\beqa
\vec{\mathcal{E}}(x^\pr)\equiv\vec{E}[\vec{x}^\pr+\vec{v}t^\pr,t^\pr+\f{\vec{x}^\pr\cdot \vec{v}}{c^2}]~,
~~\vec{\mathcal{B}}(x^\pr)\equiv\vec{B}[\vec{x}^\pr+\vec{v}t^\pr,t^\pr+\f{\vec{x}^\pr\cdot \vec{v}}{c^2}]~,
\eeqa
which, after using a new notation of variable $'z'$ to avoid confusion, can be written as
\beqa
\vec{\mathcal{E}}(z)&=&\vec{E}[\vec{z}+\vec{v} z_0,z_0+\f{\vec{v}\cdot\vec{z}}{c^2}]~,
~~\vec{\mathcal{B}}(z)=\vec{B}[\vec{z}+\vec{v} z_0,z_0+\f{\vec{v}\cdot\vec{z}}{c^2}].
\label{form:galilean}
\eeqa
This is naively the (inverse) Galilean transformed coordinate variables for electromagnetic fields if we neglect the $(\vec{v}\cdot\vec{z})/c^2$ term in the inverse transformation of $z_0$ variable.

One of Maxwell's equation at the loop-rest $\Sigma$ frame is
\beqa
\nabla^\pr\tm \vec{E}^\pr(x^\pr)&=&-\f{\pa}{\pa t^\pr} \vec{B}^\pr(x^\pr)~.
\label{loop-rest}
\eeqa
Considering eq.(\ref{huati-E-B}), the above equation can be rewritten as
\footnote{The second equation can be obtained from the first equation by simply changing variables
from $x'$ to $x$.
As different variables correspond to different reference frame, the change of variables also
corresponds to the transformations between different reference frames, as discussed in Appendix \ref{appendix:2} }
\beqa
\nabla^\pr \tm (\vec{\mathcal{E}}(x^\pr)+\vec{v}\tm \vec{\mathcal{B}}(x^\pr))=-\f{\pa}{\pa t^\pr}\(\vec{\mathcal{B}}(x^\pr)-\f{\vec{v}\tm \vec{\mathcal{E}}(x^\pr)}{c^2}\),\nn\\
\Rightarrow \nabla \tm (\vec{\mathcal{E}}(x)+\vec{v}\tm \vec{\mathcal{B}}(x))=-\f{\pa}{\pa t}\(\vec{\mathcal{B}}(x)-\f{\vec{v}\tm \vec{\mathcal{E}}(x)}{c^2}\)~.
\label{marxwell1}
\eeqa
The forms of the functions $\vec{\mathcal{E}}$ and $\vec{\mathcal{B}}$ are given by eq.(\ref{function:form}).
 Here $\nabla^\pr$ and $\nabla$ denote the derivative with respect to the $x^\pr\equiv(\vec{x}^\pr,t^\pr)$ coordinates and $x\equiv(\vec{x},t)$ coordinates, respectively. Note that the replacement in eq.(\ref{marxwell1}) is fairly non-trivial if it is to be understood as the changing of reference frame instead of simply changing of variables.
The derivative of the original $x^\pr$ coordinates should be recasted into the $x$ variables.
The $\vec{\mathcal{E}}$ and $\vec{\mathcal{B}}$ fields, which are given in terms of $\vec{E}$ and $\vec{B}$ fields with inversely Lorentz transformation coordinates, are also in the original $x^\pr$ coordinates.
New terms from the change of the derivative cancel the terms from  the change of variables for
the $\vec{\mathcal{E}}$ and $\vec{\mathcal{B}}$ fields up to the order of $v/c$. Details of the cancelation
are given in Appendix \ref{appendix:2}.

If the forms $\vec{E}(x),\vec{B}(x)$ (instead of the form of the functions $\vec{\mathcal{E}}(x)$ and $\vec{\mathcal{B}}(x)$)
are adopted, we can recover the Maxwell's equations in the laboratory $\Sigma$ frame.
For example, we can check one of the equation in Maxwell's equations
\beqa
\nabla\tm \vec{E}=-\f{\pa}{\pa t} \vec{B}.
\label{lab}
\eeqa
In the $\Sigma^\pr$ reference (loop-rest) frame, the left-handed side of eq.(\ref{loop-rest}) is given by
\beqa
&&\ga \nabla^\pr_{x^\pr}\tm\(\vec{E}(x)+\vec{v}\tm \vec{B}(x)\)\nn\\
&=&\ga^2\(\nabla+\f{\vec{v}}{c^2}\f{\pa}{\pa t}\)\tm\(\vec{E}(x)+\vec{v}\tm \vec{B}(x)\)\nn\\
&=&\ga^2\(\nabla\tm \vec{E}(x)+\f{\vec{v}}{c^2}\tm\f{\pa \vec{E}(x)}{\pa t}-(\vec{v}\cdot \nabla)\vec{B}(x)-\f{\vec{v}}{c^2}\tm(\vec{v}\tm\(\nabla\tm \vec{E}(x)\))\),
\eeqa
while the right-handed side of eq.(\ref{loop-rest}) is
\beqa
&&-\ga\f{\pa}{\pa t^\pr} \(\vec{B}(x)-\f{\vec{v}\tm \vec{E}(x)}{c^2}\)\nn\\
&=&-\ga^2\(\vec{v}\cdot \nabla+\f{\pa}{\pa t}\)\(\vec{B}(x)-\f{\vec{v}\tm \vec{E}(x)}{c^2}\)\nn\\
&=&\ga^2\(-\f{\pa }{\pa t} \vec{B}(x)-(\vec{v}\cdot \nabla)\vec{B}(x)+\f{\vec{v}}{c^2}\tm\f{\pa \vec{E}(x)}{\pa t} -\f{\vec{v}}{c^2}\tm(\vec{v}\tm\(\nabla\tm \vec{E}(x)\))\)~,
\eeqa
where
\beqa
\(\vec{v}\cdot \nabla\)(\vec{v}\tm\vec{E})&=&\nabla\[\vec{v}\cdot\(\vec{v}\tm\vec{E}\)\]-\vec{v}\tm\(\nabla\tm(\vec{v}\tm \vec{E})\)~,\nn\\
&=&-\vec{v}\tm\(\nabla\tm(\vec{v}\tm \vec{E})\),\nn\\
&=&\vec{v}\tm \[\(\vec{v}\cdot\nabla\)\vec{E}\]~,\nn\\
&=&-\vec{v}\tm\[\vec{v}\tm\(\nabla\tm \vec{E}\)\]~.
\eeqa
Comparing the above equations we just reproduce the ordinary form of Maxwell's equations in eq.(\ref{lab}).

So it is important to notify which forms of functions for electromagnetic fields,
 $\{\vec{E}(x),\vec{B}(x)\}$ or $\{\vec{\mathcal{E}}(x),\vec{\mathcal{B}}(x)\}$,
are used in the equations.
In the Hertz and extended Hertz equations, we should use the transformed function forms
$\vec{\mathcal{E}}(x)$ and $\vec{\mathcal{B}}(x)$ in eq.(\ref{function:form}).
The ordinary $\vec{E}(x)$ and $\vec{B}(x)$ should be used in the ordinary forms of Maxwell's equations.

Unless otherwise specified, the coordinates for the fields and the derivatives take the value $(\vec{x},t)$; the functions for
electromagnetic fields take the form in eq.(\ref{function:form}).

The (extended) Hertz-form Maxwell's equations can be deduced from ordinary Maxwell's equations order by order in $v$. From eq.(\ref{marxwell1}), after neglecting the $v/c$ term, we can reproduce the Hertz equation
\beqa
\nabla \tm (\vec{\mathcal{E}}+\vec{v}\tm \vec{\mathcal{B}})=-\f{\pa}{\pa t}\vec{\mathcal{B}}.
\label{result1}
\eeqa
Taking into account the $v/c$ correction, we will have an additional term
\beqa
\f{\vec{v}}{c^2}\tm \f{\pa}{\pa t} \vec{\mathcal{E}}~.
\eeqa
The expression $\pa \vec{\mathcal{E}}/\pa t$ can be deduced as
\beqa
\nabla \tm \[\vec{\mathcal{B}}-\f{\vec{v}\tm \vec{\mathcal{E}}}{c^2}\]= \mu \vec{\mathcal{J}}+ \mu\epsilon \f{\pa}{\pa t} \(\vec{\mathcal{E}}+\vec{v}\tm \vec{\mathcal{B}}\)-\mu\tl{\rho}\vec{v}~,
\label{marxwell2}
\eeqa
from one of the Maxwell's equations in the loop-rest $\Sigma'$ frame
after the substitution with coordinates-transformed electromagnetic fields and sources
\beqa
\nabla^\pr \tm \vec{B}^\pr(x^\pr)= \mu \vec{J}^\pr(x^\pr)+\mu\epsilon \f{\pa}{\pa t^\pr} \vec{E}^\pr(x^\pr)~.
\eeqa
We have
\beqa
\f{\pa}{\pa t} \vec{\mathcal{E}}= \f{1}{\mu\epsilon}\nabla \tm \(\vec{\mathcal{B}}-\f{\vec{v}\tm \vec{\mathcal{E}}}{c^2}\)-\f{1}{\epsilon} \vec{\mathcal{J}}-
 \vec{v}\tm \(\f{\pa}{\pa t}\vec{\mathcal{B}}\)+\mu\tl{\rho}\vec{v}.
\eeqa
Substituting back into the expressions of (\ref{marxwell1}), we have
\beqa
\nabla \tm (\vec{\mathcal{E}}+\vec{v}\tm \vec{\mathcal{B}})
&=&-\f{\pa}{\pa t}\vec{\mathcal{B}}+\f{\vec{v}}{c^2}\tm \f{\pa}{\pa t}\vec{\mathcal{E}}\nn\\
&\approx&-\f{\pa}{\pa t}\vec{\mathcal{B}}+\f{1}{\mu\epsilon}\f{\vec{v}}{c^2}\tm \(\nabla \tm \vec{\mathcal{B}}\) -\f{1}{\epsilon}\f{\vec{v}}{c^2}\tm \vec{\mathcal{J}}~,
\label{eqn:1}
\eeqa
after neglecting the $v^2$ terms.
We also have
\beqa
\vec{v}\tm \(\nabla \tm \vec{\mathcal{B}}\)&=& \nabla(\vec{v}\cdot\vec{\mathcal{B}})-(\vec{v}\cdot\nabla)\vec{\mathcal{B}}~\nn\\
&=&-(\vec{v}\cdot\nabla)\vec{\mathcal{B}}\nn\\
&=&\nabla \tm \(\vec{v}\tm \vec{\mathcal{B}}\),
\eeqa
for constant velocity $\vec{v}$ and $\vec{v}\cdot\vec{\mathcal{B}}=0$.

 After neglecting the source $\vec{\mathcal{J}}$ term or the source vector is parallel to $\vec{v}$, we can arrive at the coefficient of $\nabla \tm (\vec{v}\tm \vec{\mathcal{B}})$ within eq.(\ref{eqn:1})
\beqa
\nabla \tm (\vec{\mathcal{E}}+\al\vec{v}\tm \vec{\mathcal{B}})=-\f{\pa}{\pa t}\vec{\mathcal{B}},
\label{result0}
\eeqa
with
\beqa
\al=1-\f{1}{\epsilon\mu}\f{1}{c^2}=1-\f{\epsilon_0\mu_0}{\epsilon\mu}=\f{\epsilon\mu-\epsilon_0\mu_0}{\epsilon\mu}
\eeqa
and the expression for light speed in the vacuum
\beqa
c^2=\f{1}{\epsilon_0\mu_0}.
\eeqa
We see that the expression for $\al$ is just the form given in \cite{ru}.

Similarly, we can substitute the Hertz equation for $\nabla \tm \vec{\mathcal{E}}$ into eq.(\ref{marxwell2})
\beqa
\nabla \tm \[\vec{\mathcal{B}}-\f{\vec{v}\tm \vec{\mathcal{E}}}{c^2}\]= \mu \(\vec{\mathcal{J}}-\vec{v}\tl{\rho}\)+ \mu\epsilon \f{\pa}{\pa t} \(\vec{\mathcal{E}}+\vec{v}\tm \vec{\mathcal{B}}\)~,
\eeqa
to obtain
\beqa
\nabla \tm \[\vec{\mathcal{B}}-\f{\vec{v}\tm \vec{\mathcal{E}}}{c^2}\]=\mu \(\vec{\mathcal{J}}-\vec{v}\tl{\rho}\)+ \mu\epsilon \f{\pa}{\pa t} \vec{\mathcal{E}}-\mu\epsilon \vec{v}\tm \[\nabla \tm (\vec{\mathcal{E}}+\al\vec{v}\tm \vec{\mathcal{B}})\],
\eeqa
with
\beqa
\f{\pa}{\pa t}\(\vec{v}\tm \vec{\mathcal{B}}\)=\vec{v}\tm \f{\pa }{\pa t}\vec{\mathcal{B}}=-\vec{v}\tm \[\nabla \tm (\vec{\mathcal{E}}+\al\vec{v}\tm \vec{\mathcal{B}})\]~.
\eeqa
Again, after neglecting the $v^2$ term, we arrive at
\beqa
\nabla \tm \[\vec{\mathcal{B}}-\f{\vec{v}\tm \vec{\mathcal{E}}}{c^2}\]= \mu \(\vec{\mathcal{J}}-\vec{v}\tl{\rho}\)+ \mu\epsilon \f{\pa}{\pa t} \vec{\mathcal{E}}-\mu\epsilon \vec{v}\tm \(\nabla \tm \vec{\mathcal{E}}\).
\eeqa
Using
\beqa
\vec{v}\tm \(\nabla \tm \vec{\mathcal{E}}\)&=& \nabla(\vec{v}\cdot\vec{\mathcal{E}})-(\vec{v}\cdot\nabla)\vec{\mathcal{E}}\nn\\
&=&-(\vec{v}\cdot\nabla)\vec{\mathcal{E}}\nn\\
&=&\nabla \tm \(\vec{v}\tm \vec{\mathcal{E}}\)-(\nabla\cdot \vec{\mathcal{E}}) \vec{v} \nn\\
&=&\nabla \tm \(\vec{v}\tm \vec{\mathcal{E}}\)-\f{1}{\epsilon}\tl{\rho} \vec{v}~,
\eeqa
we can arrive at
\beqa
\nabla \tm \[\vec{\mathcal{B}}-\mu_0\epsilon_0{\vec{v}\tm \vec{\mathcal{E}}}\]= \mu \vec{\mathcal{J}} + \mu\epsilon \f{\pa}{\pa t} \vec{\mathcal{E}}-\mu\epsilon \nabla\tm \( \vec{v}\tm \vec{\mathcal{E}}\),
\label{result2}
\eeqa
which is just the form
\beqa
\nabla \tm \[\vec{\mathcal{B}}+\mu\epsilon \alpha{\vec{v}\tm \vec{\mathcal{E}}}\]&=& \mu \vec{\mathcal{J}}+ \mu\epsilon \f{\pa}{\pa t} \vec{\mathcal{E}}~,\nn\\
\nabla \tm \[\vec{\mathcal{H}}+ \alpha{\vec{v}\tm \vec{\mathcal{D}}}\]&=& \vec{\mathcal{J}}+ \f{\pa}{\pa t} \vec{\mathcal{D}},
\eeqa
with
\beqa
\alpha\equiv\f{\epsilon\mu-\epsilon_0\mu_0}{\epsilon\mu}~.
\eeqa
So the generalization of Hertz-form equations to include $\alpha$ can be seen to come from the non-relativisitic
expansion
of Lorentz-transformed Maxwell's equations, using an alternative form of functions for electromagnetic fields.

Other extended Herz-form Maxwell equations can be obtained in a similar way
\beqa
&&\nabla\cdot\({\vec{\mathcal{E}}+\vec{v}\tm\vec{\mathcal{B}}}\)=\f{1}{\epsilon}\(\tl{\rho}-\f{\vec{v}\cdot\vec{\mathcal{J}}}{c^2}\)\nn\\
\Rightarrow&&\nabla\cdot{\vec{\mathcal{E}}}-\vec{v}\cdot\(\nabla\tm\vec{\mathcal{B}}\)=\f{1}{\epsilon}\(\tl{\rho}-\f{\vec{v}\cdot\vec{\mathcal{J}}}{c^2}\)\nn\\
\Rightarrow&&\nabla\cdot{\vec{\mathcal{E}}}-\vec{v}\cdot\(\mu \vec{\mathcal{J}}+ \f{1}{c^2} \f{\pa}{\pa t} \vec{\mathcal{E}}+\alpha(\vec{v}\cdot\nabla){\vec{\mathcal{E}}} \)=\f{1}{\epsilon}\tl{\rho}-\mu\vec{v}\cdot\vec{\mathcal{J}}\nn\\
\Rightarrow&&\nabla\cdot{\vec{\mathcal{E}}}-\f{\vec{v}}{c^2}\cdot\( \f{\pa}{\pa t} \vec{\mathcal{E}}+\alpha(\vec{v}\cdot\nabla){\vec{\mathcal{E}}} \)=\f{1}{\epsilon}\tl{\rho}\nn\\
\Rightarrow&&\nabla\cdot{\vec{\mathcal{E}}}=\f{1}{\epsilon}\tl{\rho}~,
\label{eqn:div1}
\eeqa
with
\beqa
\f{d}{dt}(\vec{v}\cdot{\vec{\mathcal{E}}})\equiv 0
\Rightarrow \vec{v}\cdot\f{\pa}{\pa t}{\vec{\mathcal{E}}}=0.
\eeqa
We neglect the order $v^2$ term in the fourth line of eq.(\ref{eqn:div1}) in the last step.

We also have
\beqa
&&\nabla\cdot\(\vec{\mathcal{B}}-\f{\vec{v}\tm\vec{\mathcal{E}}}{c^2}\)=0~,\nn\\
\Rightarrow&&\nabla\cdot{\vec{\mathcal{B}}}-\f{\vec{v}}{c^2}\cdot\(\nabla\tm\vec{\mathcal{E}}\)=0~,\nn\\
\Rightarrow&&\nabla\cdot{\vec{\mathcal{B}}}-\f{\vec{v}}{c^2}\cdot\(-\f{\pa}{\pa t}\vec{\mathcal{B}}+\al \(\vec{v}\cdot\nabla\)\vec{\mathcal{B}}\)=0~,\nn\\
\Rightarrow&&\nabla\cdot{\vec{\mathcal{B}}}=0~,\nn\\
\eeqa
with
\beqa
\f{d}{dt}(\vec{v}\cdot{\vec{\mathcal{B}}})\equiv 0
\Rightarrow \vec{v}\cdot\f{\pa}{\pa t}{\vec{\mathcal{B}}}=0~,\nn
\eeqa
after neglecting the order $v^2$ term.

The new extended Hertz-form equations can be written together as
\beqa
\nabla\cdot \vec{\mathcal{E}}&=& \f{1}{\epsilon}\tl{\rho}~,~~\\
\nabla\cdot \vec{\mathcal{B}}&=& 0~,\\
\nabla \tm \vec{\mathcal{B}}&=&\mu \vec{\mathcal{J}}+ \mu\epsilon \f{\pa}{\pa t} \vec{\mathcal{E}}+\mu\epsilon \alpha \(\vec{v}\cdot\nabla\)\vec{\mathcal{E}}~,~\label{eqn:hertz3}\\
\nabla \tm \vec{\mathcal{E}}&=& -\f{\pa}{\pa t} \vec{\mathcal{B}}+\alpha \(\vec{v}\cdot\nabla\)\vec{\mathcal{B}}~\label{eqn:hertz4}.
\eeqa

In the case of conductors with $\mu,\epsilon\ra \infty$, the value of $\alpha$ tends to $1$. We should note that, even when $\al\ra 1$, our new extended Hertz-form Maxwell's equations can not recover the original Hertz-form Maxwell's equations. In the original Hertz-form Maxwell's equations, the $\pa t$ derivative always accompany with the $\vec{v}\cdot\nabla$ term so as that they can be combined into a $D_t$ derivative
\beqa
D_t\equiv \f{\pa}{\pa t}-(\vec{v}\cdot\nabla)~.
\eeqa
However, in our new (extended) Hertz-form Maxwell's equations, there is a sign difference for the $\alpha$ term in
eq.(\ref{eqn:hertz3}), which can not be written as $D_t$ in the $\al\ra 1$ limit. Using the $-\vec{v}$ instead
of $\vec{v}$ can not change both eq.(\ref{eqn:hertz3}) and eq.(\ref{eqn:hertz4}) into the $D_t$ form in the
$\al\ra 1$ limit due to the relative sign difference in $\pa_t$ for the two equations.
Such new extended Hertz-form Maxwell's equations can not recover the Galilean transformed Maxwell's
equations (the original Hertz-form Maxwell's equations) in the $\al\ra 1$ limit.

So, it is obvious that the extended Hertz-form equations can be deduced from the non-relativistic expansion of
Lorentz transformation of Maxwell's equations, which adopt an alternative form of functions for electromagnetic fields.
Although the original Hertz-form equations preserve Galilean invariance, the extend Hertz-form equations lose such
an invariance.  As the propagating electromagnetic waves are intrinsically relativistic, the extended Hertz form
(from non-relativistic expansion) should be used when the electromagnetic fields can be seen as the background.

It is worth noting that the $\{\vec{\mathcal{E}},\vec{\mathcal{B}}\}$ set description for electromagnetic field is an equivalent traveling wave form description for electromagnetic field. Such a traveling wave form description satisfies the extended Hertz-form Maxwell's equations. That is, the solution of extended Hertz-form Maxwell's equations describe the traveling wave form electromagnetic field.

\section{\label{sec-3}Discussions about the superluminal behavior}
From the previous extended Hertz equations involving $\al$, we can deduce the D'Alembert equation
for $\vec{\mathcal{E}}$ (and $\vec{\mathcal{B}}$). For simply, we consider only the plane wave solutions
with
\beqa
\vec{\mathcal{E}}(x)=\vec{\mathcal{E}}_0 \exp\[i \(\vec{k}\cdot\vec{x}-\omega t\)\],~~
\vec{\mathcal{B}}(x)=\vec{\mathcal{B}}_0 \exp\[i \(\vec{k}\cdot\vec{x}-\omega t\)\].
\eeqa
The extended Hertz equations without sources can be rewritten as
\beqa
\nabla\cdot \vec{\mathcal{E}}&=& 0~,~~~\nabla\cdot \vec{\mathcal{B}}=0~,\nn\\
\nabla \tm \vec{\mathcal{B}}&=& \mu\epsilon \f{\pa}{\pa t} \vec{\mathcal{E}}+\mu\epsilon \alpha \(\vec{v}\cdot\nabla\)\vec{\mathcal{E}}~,~\nn\\
\nabla \tm \vec{\mathcal{E}}&=& -\f{\pa}{\pa t} \vec{\mathcal{B}}+\alpha \(\vec{v}\cdot\nabla\)\vec{\mathcal{B}}~.
\eeqa
Take the \textit{curl} operation for the fourth equation, we can obtain
\beqa
-\nabla^2 \vec{\mathcal{E}}&=&-\f{\pa}{\pa t}\[\mu\epsilon \f{\pa}{\pa t} \vec{\mathcal{E}}+\mu\epsilon \alpha \(\vec{v}\cdot\nabla\)\vec{\mathcal{E}}\]+\alpha  \nabla\tm \[\(\vec{v}\cdot\nabla\)\vec{\mathcal{B}}\]~,\nn\\
&=&-\f{1}{\tl{c}^2}\[\f{\pa^2}{\pa t^2}\vec{\mathcal{E}}+i\al\(\vec{v}\cdot\vec{k}\)\f{\pa}{\pa t} \vec{\mathcal{E}}\]
+i\alpha \(\vec{v}\cdot\vec{k}\)\[\nabla\tm \vec{\mathcal{B}}\]~,\nn\\
&=&-\f{1}{\tl{c}^2}\[\f{\pa^2}{\pa t^2}\vec{\mathcal{E}}+i\al\(\vec{v}\cdot\vec{k}\)\f{\pa}{\pa t}\vec{\mathcal{E}}\]
+i\alpha \(\vec{v}\cdot\vec{k}\)\f{1}{\tl{c}^2}\[\f{\pa}{\pa t} +i\al\(\vec{v}\cdot\vec{k}\)\]\vec{\mathcal{E}}~,\nn\\
&=&-\f{1}{\tl{c}^2}\f{\pa^2}{\pa t^2}\vec{\mathcal{E}}-\f{\al^2}{\tl{c}^2}\(\vec{v}\cdot\vec{k}\)^2\vec{\mathcal{E}}~.
\eeqa
with the light speed in the media $1/\tl{c}^2=\mu\epsilon$.

So the resulting D'Alembert equation can be written as
\beqa
\[\nabla^2-\f{1}{\tl{c}^2}\f{\pa^2}{\pa t^2}-\f{\al^2}{\tl{c}^2}\(\vec{v}\cdot\vec{k}\)^2\] \vec{\mathcal{E}}=0~.
\eeqa
We have the following discussions:
\bit
\item $\vec{v}\cdot\vec{k}=0$, that is, the direction $\vec{v}$ is perpendicular to propagating direction $\vec{k}$.
   The D'Alembert equation takes the ordinary form with the light speed $\tl{c}^2=1/(\mu\epsilon)$.

\item $\vec{v}\parallel\vec{k}$. The D'Alembert equation takes the form
\beqa
\[\nabla^2-\f{1}{\tl{c}^2}\f{\pa^2}{\pa t^2}-\f{\al^2}{\tl{c}^2} v^2 k^2\] \vec{\mathcal{E}}=0~.
\eeqa
In momentum space, we have
\beqa
-k^2+\f{\omega^2}{\tl{c}^2}-\f{\al^2}{\tl{c}^2} v^2 k^2=0~.
\eeqa
So
\beqa
\(1+\f{\al^2}{\tl{c}^2}v^2\)k^2=\f{\omega^2}{\tl{c}^2}~.
\eeqa
We can define a new quantity, {\it apparent light speed} $\tl{c}^\pr$, to satisfy
\beqa
\tl{c}^{\pr 2}=\tl{c}^2+\al^2v^2~,
\eeqa
so as that the ordinary relation $k^2\tl{c}^{\pr 2}=\omega^2$ is preserved for massless photon.
It is obvious that the quantity $\tl{c}^\pr$ is always larger than the light speed in the media.

\eit
 From previous discussions, although it is not problematic for the apparent light speed $\tl{c}^\pr$ of the electromagnetic field
to be larger than the ordinary light speed in the media,  the quantity $\tl{c}^\pr$ apparently exceed the ordinary vacuum light speed if this D'Alembert equation is applied for the vacuum. However, it is interesting to note that the value $\al=0$ is hold for vacuum. So, the value of apparent light speed in the vacuum  is still equal to the vacuum light speed. Even though $\tl{c}^{\pr 2}=c^2$ holds in the vacuum, it is still possible that the apparent light speed $\tl{c}^\pr$ in the media is larger than the vacuum light speed, depending on the choice of $\al$ and $v$.

One may worry that the superluminal behavior may cause inconsistency. However, we should note that the previous
expressions in eq.(\ref{result0}) and eq.(\ref{result2}) are just the non-relativistic approximation of
Lorentz transformations for Maxwell's equations. They should not apply to the propagation of electromagnetic
wave, which is intrinsically relativistic. By going to the relativistic region, the sub-leading terms in
$v/c$ expansion should be included. The ${\al}^2k^2v^2$ term will be canceled by the
inclusion of all the subleading contributions (in $v/c$ expansion) to keep the vacuum light speed as a
constant, which is just one of the two ansatzes of special relativity. In the case that the electromagentic
field is a background field, we need not worry about the apparent superluminal behavior illustrated by
the D'Alembert equation.

\section{\label{conclusion}Conclusions}
We showed explicitly that the Hertz-form Maxwell's equations and their extensions
can be obtained from the non-relativistic expansion of Lorentz transformation of Maxwell's equations.
The explicit expression for the parameter $\alpha$ in the extended Hertz-form equations can be derived
from such a non-relativistic expansion. The extended Hertz-form equations,
which do not preserve Galilean invariance, origin from Lorentz transformation of Maxwell's equations.
Especially, the $\alpha$ parameter is of relativistic origin. The superluminal behavior illustrated by
the D'Alembert equation from the extended Hertz-form equations should be removed by including
all subleading contributions in the $v/c$ expansion, although such a superluminal behavior will not occur in the vacuum because $\al=0$. In the case that the electromagnetic
field is a background field, we need not worry about the apparent superluminal behavior of the
D'Alembert equation. We should note that in the new Hertz form and extended Hertz form equations,
the electromagnetic fields should take the forms
$ \vec{\mathcal{E}}(x)=\vec{E}(\Lambda^{-1}x)$ and $ \vec{\mathcal{B}}(x)=\vec{B}(\Lambda^{-1}x)$
while the derivatives in the equations are taken with respect to $x$.
Such a choice of description for the fields is different from the ordinary one with $\vec{E}(x)$
and $\vec{B}(x)$, which are well known to satisfy the ordinary Maxwell's equations.
The descriptions of electromagnetic phenomena using the function
set $\{\vec{\mathcal{E}}(x),\vec{\mathcal{B}}(x)\}$ and the function set $(\vec{E}(x),\vec{B}(x))$ are equivalent,
with the $\{\vec{\mathcal{E}}(x),\vec{\mathcal{B}}(x)\}$ description satisfying
the new extended Hertz-form  Maxwell's equations in the low speed approximation. The solution of 
(extended) Hertz-form Maxwell's equations describe the traveling wave form electromagnetic field.

We should mention that ordinary magnetic monopole extension of Maxwell's equations use the form of
functions $\vec{E}(x)$ and $\vec{B}(x)$. If we adopt the $\vec{\mathcal{E}}(x)$ and $\vec{\mathcal{B}}(x)$
descriptions of the electromagnetic fields, we can also extend the (extended) Hertz-form Maxwell's equations
with additional topological terms.

Besides, as the (extended) Hertz equations can be readily deduced from special relativity,
the aether assumption is no-longer needed here. It is just the non-relativistic limit description
of Maxwell's equations with an alternative set of functions for electromagnetic fields,
although they can approximately preserve Galilean invariance.

\acknowledgments
We are very grateful to Rong-Gen Cai, Ming Yu, Qun Wang and Qing Wang for reading our draft carefully
and giving good suggestions.  We thank Wei-Mou Zheng and Zhong-Lin Wang for discussions.

\appendix
\section{\label{appendix:2}The non-relativistic expansion of typical terms in the Lorentz-transformed Maxwell's equations}
In eq.(\ref{marxwell1}), the expressions in both coordinates are unchanged. It can surely be seen as the naive
change of variables. It can also be seen physically by directly adopting the transformations of reference frame.

The expansion of electromagnetic fields in $x^\pr$ variable can be expressed in terms of $x$ variable as
\beqa
\vec{\mathcal{E}}(\vec{x}^\pr,t^\pr)\approx \vec{\mathcal{E}}(\vec{x}- t\vec{v},t-\f{\vec{v}\cdot \vec{x}}{c^2})\approx \vec{\mathcal{E}}(\vec{x},t)-t \(\vec{v}\cdot \nabla\) \vec{\mathcal{E}}(\vec{x},t)-\f{\vec{v}\cdot \vec{x}}{c^2}\f{\pa}{\pa t} \vec{\mathcal{E}}(\vec{x},t),
\eeqa
for small $v$. Similarly expansion can be given for other functions in the Hertz and extended Hertz equations.
We keep only the leading order of $v$ expansion without taking into account the $\ga$ factor,
which will give higher order terms of $v^2/c^2$ in the expansion.
Firstly, we consider the transformation
\beqa
\nabla\tm \vec{\mathcal{B}}(\vec{x},t)\ra \nabla^\pr \tm \vec{\mathcal{B}}(\vec{x}^\pr,t^\pr)~.
\eeqa
From the Lorentz transformation law of $\pa_\mu$, we can obtain the expressions for $\nabla$
\beqa
\nabla^\pr &\ra& \ga\(\nabla+\f{\vec{v}}{c^2}\f{\pa}{\pa t}\),\nn\\
\f{\pa}{\pa t^\pr} &\ra&\ga\({\vec{v}}\cdot \nabla+ \f{\pa}{\pa t}\).\nn\\
\eeqa
So we have
\beqa
\nabla^\pr\tm \vec{\mathcal{E}}(\vec{x}^\pr,t^\pr)&\approx&\ga\(\nabla+\f{\vec{v}}{c^2}\f{\pa}{\pa t}\)\tm\[ \vec{\mathcal{E}}(\vec{x},t)-t \(\vec{v}\cdot \nabla\) \vec{\mathcal{E}}(\vec{x},t)-\f{\vec{v}\cdot \vec{x}}{c^2}\f{\pa}{\pa t} \vec{\mathcal{E}}(\vec{x},t)\]\nn\\
&\approx& \nabla \tm \vec{\mathcal{E}}(\vec{x},t)+\f{\vec{v}}{c^2}\f{\pa}{\pa t}\tm \vec{\mathcal{E}}(\vec{x},t)-\f{1}{c^2}\nabla\tm \[(\vec{v}\cdot \vec{x}) \f{\pa}{\pa t} \vec{\mathcal{E}}(\vec{x},t) \]
-t\nabla\tm \[\(\vec{v}\cdot \nabla \)\vec{\mathcal{E}}(\vec{x},t)\]~ \nn\\
&\approx& \nabla \tm \vec{\mathcal{E}}(\vec{x},t)+\f{1}{\mu\epsilon}\f{\vec{v}}{c^2}\tm\[\nabla\tm \vec{\mathcal{B}}(\vec{x},t)\]-\f{1}{\mu\epsilon}\f{\vec{v}}{c^2}\tm\[\nabla\tm \vec{\mathcal{B}}(\vec{x},t)\]-t\nabla\tm \[\(\vec{v}\cdot \nabla\) \vec{\mathcal{E}}(\vec{x},t)\]\nn\\
&=& \nabla \tm \vec{\mathcal{E}}(\vec{x},t)-t\nabla\tm \[\(\vec{v}\cdot \nabla\) \vec{\mathcal{E}}(\vec{x},t)\]~,
\label{eqn:3}
\eeqa
with
\beqa
\nabla\tm \[(\vec{v}\cdot \vec{x}) \f{\pa}{\pa t} \vec{\mathcal{E}}(\vec{x},t)\]&=&\nabla(\vec{v}\cdot \vec{x})\tm \f{\pa}{\pa t} \vec{\mathcal{E}}(\vec{x},t)+(\vec{v}\cdot \vec{x}) \f{\pa}{\pa t}\(\nabla\tm \vec{\mathcal{E}}(\vec{x},t)\)\nn\\
&=& \f{1}{\mu\epsilon}\vec{v}\tm (\nabla\tm \vec{\mathcal{B}}-\f{1}{\mu}\vec{\mathcal{J}})-(\vec{v}\cdot \vec{x})\f{\pa^2}{\pa t^2}\vec{\mathcal{B}}~.
\eeqa
The term involving $\pa^2\vec{\mathcal{B}}/\pa t^2$ is $1/c^2$ suppressed in comparison with the first term as $c^2=1/(\mu\epsilon)$ and $\vec{\mathcal{J}}$ always parallel to $\vec{v}$. So we can neglect such terms.
Note that the cancelation of the $\pa \vec{\mathcal{E}}/\pa t$ terms in the second line of eq.(\ref{eqn:3}) is automatic, which do not depend on the
explicit form of $\pa \vec{\mathcal{E}}/\pa t$ in Maxwell's equations.

Up to $v/c$ order, we have
\beqa
\nabla^\pr\tm \(\vec{v}\tm \vec{\mathcal{B}}(\vec{x}^\pr,t^\pr)\)\approx \nabla \tm \(\vec{v}\tm \vec{\mathcal{B}}(\vec{x},t)\)+o(v^2/c^2)~.
\eeqa
The right-handed side of eq.(\ref{marxwell1}) is
\beqa
\f{\pa}{\pa t^\pr} \vec{\mathcal{B}}(\vec{x}^\pr,t^\pr)&\approx &\({\vec{v}}\cdot \nabla+ \f{\pa}{\pa t}\)\[ \vec{\mathcal{B}}(\vec{x},t)-t \(\vec{v}\cdot \nabla\) \vec{\mathcal{B}}(\vec{x},t)\]\nn\\
&\approx& \(\vec{v}\cdot \nabla\)\vec{\mathcal{B}}({x})+\f{\pa}{\pa t}\vec{\mathcal{B}}({x})-\({\vec{v}}\cdot \nabla\)\vec{\mathcal{B}}({x})-t\f{\pa}{\pa t}\(\(\vec{v}\cdot \nabla\) \vec{\mathcal{B}}({x})\)\nn\\
&=&\f{\pa}{\pa t}\vec{\mathcal{B}}(\vec{x},t)-t\f{\pa}{\pa t}\(\nabla\tm\(\vec{v} \tm\vec{\mathcal{B}}({x})\)\)\nn\\
&=&\f{\pa}{\pa t}\vec{\mathcal{B}}(\vec{x},t)-t\(\nabla\tm\(\vec{v} \tm \f{\pa}{\pa t}\vec{\mathcal{B}}({x})\)\)\nn\\
&\approx&\f{\pa}{\pa t}\vec{\mathcal{B}}(\vec{x},t)+t\(\nabla\tm\(\vec{v} \tm \(\nabla\tm\vec{\mathcal{E}}({x})\)\)\)\nn\\
&=&\f{\pa}{\pa t}\vec{\mathcal{B}}(\vec{x},t)+t\(\nabla\tm\(\nabla(\vec{v}\cdot\vec{\mathcal{E}})
-(\vec{v}\cdot\nabla)\vec{\mathcal{E}}({x})\)\)~,\nn\\
&=&\f{\pa}{\pa t}\vec{\mathcal{B}}(\vec{x},t)-t\(\nabla\tm\(\vec{v}\cdot\nabla\)\vec{\mathcal{E}}({x})\)~,\nn\\
\eeqa
and
\beqa
\f{\pa}{\pa t^\pr}\[ \f{\vec{v}}{c^2}\tm \vec{\mathcal{E}}(\vec{x}^\pr)\]
&\approx& \f{\pa}{\pa t}\[ \f{\vec{v}}{c^2}\tm \vec{\mathcal{E}}(\vec{x})\]~,
\eeqa
up to $o(v)$.

So the Maxwell's equation
\beqa
\nabla^\pr\tm \vec{E}^\pr(x^\pr)&=&-\f{\pa}{\pa t^\pr} \vec{B}^\pr(x^\pr)\nn\\
\Rightarrow \nabla^\pr\tm (\vec{\mathcal{E}}+\vec{v}\tm \vec{\mathcal{B}})(x^\pr)
&=&-\f{\pa}{\pa t^\pr} \(\vec{\mathcal{B}}-\f{\vec{v}\tm \vec{\mathcal{E}}}{c^2}\)(x^\pr)~,
\eeqa
can be recast into the form
\beqa
\nabla \tm \vec{\mathcal{E}}(x)+\nabla \tm \(\vec{v}\tm \vec{\mathcal{B}}(x)\)
&=&-\f{\pa}{\pa t}\vec{\mathcal{B}}(x)+\f{\vec{v}}{c^2} \tm \f{\pa}{\pa t}\(\vec{\mathcal{E}}(x)\)~.
\eeqa

The extension to the formula involving $\nabla\tm \vec{B}$ is straightforward. The divergence related expression in the Maxwell's equations is transformed as
\beqa
\nabla^\pr \cdot \vec{\mathcal{E}}(x^\pr)&\approx&(\nabla+\f{\vec{v}}{c^2}\f{\pa}{\pa t})\cdot \[\vec{\mathcal{E}}(\vec{x},t)-t \(\vec{v}\cdot \nabla\) \vec{\mathcal{E}}(\vec{x},t)-\f{\vec{v}\cdot \vec{x}}{c^2}\f{\pa}{\pa t} \vec{\mathcal{E}}(\vec{x},t)\]\nn\\
&\approx&\nabla\cdot \vec{\mathcal{E}}(\vec{x},t)+\f{\vec{v}}{c^2}\cdot\f{\pa}{\pa t}\vec{\mathcal{E}}(\vec{x},t)
-\f{1}{c^2}\[\nabla(\vec{v}\cdot \vec{x})\]\cdot\f{\pa}{\pa t} \vec{\mathcal{E}}(\vec{x},t)-\f{\vec{v}\cdot \vec{x}}{c^2}\nabla\cdot\[\f{1}{\mu\epsilon}\nabla\tm \vec{\mathcal{B}}-\f{1}{\epsilon}\vec{\mathcal{J}}\]\nn\\
&\approx&\nabla\cdot \vec{\mathcal{E}}(\vec{x},t)+\f{\vec{v}}{c^2}\cdot\f{\pa}{\pa t}\vec{\mathcal{E}}(\vec{x},t)
-\f{\vec{v}}{c^2}\cdot\f{\pa}{\pa t} \vec{\mathcal{E}}(\vec{x},t)+\f{(\vec{v}\cdot \vec{x})}{\epsilon c^2}\nabla\cdot\vec{\mathcal{J}}\nn\\
&\approx&\nabla\cdot \vec{\mathcal{E}}(\vec{x},t)+\f{(\vec{v}\cdot \vec{x})}{\epsilon c^2}\nabla\cdot\vec{\mathcal{J}}~,
\eeqa
by using
\beqa
\nabla\cdot\[(\vec{v}\cdot \vec{x})\f{\pa}{\pa t} \vec{\mathcal{E}}(\vec{x},t)\]
&=&\[\nabla({\vec{v}\cdot \vec{x}})\]\cdot \f{\pa}{\pa t} \vec{\mathcal{E}}(\vec{x},t)
+({\vec{v}\cdot \vec{x}})\nabla\cdot\f{\pa}{\pa t} \vec{\mathcal{E}}(\vec{x},t)~,\nn\\
\nabla\cdot((\vec{v}\cdot\nabla)\vec{\mathcal{E}}(\vec{x},t))
&=&-\nabla\cdot\(\nabla\tm (\vec{v}\tm\vec{\mathcal{E}}(\vec{x},t))\)=0~,
\eeqa
and
\beqa
\nabla\cdot(\nabla\tm \vec{\mathcal{B}})=0~,
\eeqa
up to $o(v)$.

Obviously, the electric density changes as
\beqa
\tl{\rho}(x^\pr)&\approx& \tl{\rho}(\vec{x},t)-t \(\vec{v}\cdot \nabla\) \rho(\vec{x},t)-\f{\vec{v}\cdot \vec{x}}{c^2}\f{\pa}{\pa t} \tl{\rho}(\vec{x},t).
\eeqa
So, the equation
\beqa
\nabla^\pr\cdot\vec{\mathcal{D}}^\pr(x^\pr)=\rho^\pr(x^\pr)~,
\eeqa
will lead to
\beqa
\nabla \cdot \vec{\mathcal{D}}(x)=\rho^\pr(x)~,
\eeqa
with the application of the conservation equation
\beqa
\f{\pa}{\pa t} \rho + \nabla\cdot \vec{J}=0~.
\eeqa
\section{\label{appendix:3} The extended Hertz equations with general choice of $\vec{v}$ direction}
In the deduction of extended Hertz equations in the main text, the assumption that $\vec{v}$ is perpendicular to both $\vec{E}$ and $\vec{B}$ is used. In this section, we will generalize the previous results to the general case with arbitrary choice of $\vec{v}$ direction.

The Lorentz transformation of electromagnetic fields are given as
\beqa
\vec{E}^\pr(x^\pr)&=&\ga(\vec{E}_{\perp}+\vec{v}\tm\vec{B})+\vec{E}_{\shortparallel}
\approx \vec{E}+\vec{v}\tm\vec{B}~,\nn\\
\vec{B}^\pr(x^\pr)&=&\ga(\vec{B}_{\perp}-\f{\vec{v}}{c^2}\tm\vec{E})+\vec{B}_{\shortparallel}
\approx\vec{B}-\f{\vec{v}}{c^2}\tm\vec{E}~,
\eeqa
at order ${\cal O}(v)$ in the $v$ expansion with the Lorentz transformation of the coordinates
\beqa
 x'=\Lambda x~.
\eeqa
Following the discussion after eq (\ref{huati-E-B}), such transformation can be rewritten with the same coordinate variables for the involved fields as
\beqa
\vec{E}^\pr&=&
\ga(\vec{\mathcal{E}}_{\perp}+\vec{v}\tm\vec{\mathcal{B}})+\vec{\mathcal{E}}_{\shortparallel}
\approx\vec{\mathcal{E}}+\vec{v}\tm\vec{\mathcal{B}} ~,\nn\\
\vec{B}^\pr&=&\ga(\vec{\mathcal{B}}_{\perp}-\f{\vec{v}}{c^2}\tm\vec{\mathcal{E}})+\vec{\mathcal{B}}_{\shortparallel}
\approx \vec{\mathcal{B}}-\f{\vec{v}}{c^2}\tm\vec{\mathcal{E}}~,
\eeqa
where
\beqa
\vec{\mathcal{E}}(x^\pr)\equiv\vec{E}[x(x^\pr)]~,~\vec{\mathcal{B}}(x^\pr)\equiv\vec{B}[x(x^\pr)]~,
\label{function:form2}
\eeqa
with
\beqa
x(x')=\Lambda^{-1} x^\pr~,
\eeqa
being the inverse coordinate transformation.

Following the approaches in the main text, it is easy to deduce the Hertz form Maxwell's equations from ordinary Maxwell's equations order by order in $v$, which gives
\beqa
\nabla\tm \vec{E}&=&-\f{\pa}{\pa t}\vec{B}~,\nn\\
\Longrightarrow\nabla \tm (\vec{\mathcal{E}}+\vec{v}\tm \vec{\mathcal{B}})&=&-\f{\pa}{\pa t}\vec{\mathcal{B}}+\f{v}{c^2}\tm \f{\pa}{\pa t}\vec{\mathcal{E}}~,\nn\\
&\approx&-\f{\pa}{\pa t}\vec{\mathcal{B}}+\f{1}{\mu\epsilon}\f{\vec{v}}{c^2}\tm \(\nabla \tm \vec{\mathcal{B}}\) -\f{1}{\epsilon}\f{\vec{v}}{c^2}\tm \vec{\mathcal{J}}~,
\label{result1}
\eeqa
After rearranging terms and neglect ${\cal O}(v^2)$ terms, we can obtain
\beqa
\nabla \tm (\vec{\mathcal{E}}+\al\vec{v}\tm \vec{\mathcal{B}})=-\f{\pa}{\pa t}\vec{\mathcal{B}}-\f{1}{\epsilon}\f{\vec{v}}{c^2}\tm \vec{\mathcal{J}}+\f{1}{\mu\epsilon}\f{1}{c^2}\nabla(\vec{v}\cdot\vec{\mathcal{B}}).
\eeqa
Similarly, replacing the new equivalent form (\ref{function:form2}) into
\beqa
\nabla\tm \vec{B}=\vec{J}+\mu\epsilon\f{\pa}{\pa t}\vec{E}~,
\eeqa
and neglecting the ${\cal O}(v^2)$ terms, we have 
\beqa
\nabla \tm \[\vec{\mathcal{B}}+\mu\epsilon \alpha{\vec{v}\tm \vec{\mathcal{E}}}\]&=& \mu \vec{\mathcal{J}}+ \mu\epsilon \f{\pa}{\pa t} \vec{\mathcal{E}}-\mu\epsilon\nabla(\vec{v}\cdot\vec{\mathcal{E}})~.
\eeqa
The remaining extended Hertz-form Maxwell's equations can be readily obtained
\beqa
\nabla\cdot{\vec{\mathcal{E}}}&=&\f{1}{\epsilon}\tl{\rho}+\f{1}{c^2}\f{\pa}{\pa t}\(\vec{v}\cdot\vec{\mathcal{E}}\)~,
\label{nabla:dotE}
\eeqa
and
\beqa
\nabla\cdot{\vec{\mathcal{B}}}&=&-\f{1}{c^2}\f{\pa}{\pa t}\(\vec{v}\cdot\vec{\mathcal{B}}\)~.
\label{nabla:dotB}
\eeqa
We should note that the sign in front of $\f{\pa}{\pa t}$ within eq.(\ref{nabla:dotB}) is different to that within eq.(\ref{nabla:dotE}), which can not be recast into the form of $\tl{\nabla}$ derivative defined by
\beqa
\tl{\nabla}\equiv \nabla-\f{\vec{v}}{c^2}\f{\pa}{\pa t}~,
\eeqa
at the same time for the two equations. That is, the low speed approximation for the traveling wave form description of electromagnetic field can not reduce to the Galilean transformed form description of electromagnetic field. That is, the low speed limit of Lorentz transformation for electromagnetic field is not the Galilean transformation.

The extended Hertz-form Maxwell's equations can be written together as
\beqa
\nabla\cdot \vec{\mathcal{E}}&=& \f{1}{\epsilon}\tl{\rho}+\f{1}{c^2}\f{\pa}{\pa t}\(\vec{v}\cdot\vec{\mathcal{E}}\)~,~~\\
\nabla\cdot \vec{\mathcal{B}}&=&-\f{1}{c^2}\f{\pa}{\pa t}\(\vec{v}\cdot\vec{\mathcal{B}}\)~,\\\nabla \tm (\vec{\mathcal{E}}+\al\vec{v}\tm \vec{\mathcal{B}})&=&-\f{\pa}{\pa t}\vec{\mathcal{B}}-\f{1}{\epsilon}\f{\vec{v}}{c^2}\tm \vec{\mathcal{J}}+\f{1}{\mu\epsilon}\f{1}{c^2}\nabla(\vec{v}\cdot\vec{\mathcal{B}})~,\\
\nabla \tm \[\vec{\mathcal{B}}+\mu\epsilon \alpha{\vec{v}\tm \vec{\mathcal{E}}}\]&=& \mu\epsilon \f{\pa}{\pa t} \vec{\mathcal{E}}+\mu \vec{\mathcal{J}} -\mu\epsilon\nabla(\vec{v}\cdot\vec{\mathcal{E}})~,
\eeqa
with
\beqa
\al=\f{\epsilon\mu-\epsilon_0\mu_0}{\epsilon\mu}~.\nn
\eeqa

\end{document}